# The Potential of Using Google Expeditions and Google Lens Tools under STEM-education in Ukraine

Yevhenii B. Shapovalov[0000-0003-3732-9486], Zhanna I. Bilyk[0000-0002-2092-5241], Artem I. Atamas[0000-0002-8709-3208] and Viktor B. Shapovalov[0000-0001-6315-649X]

National Center "Junior Academy of Sciences of Ukraine",
38/44, Dehtiarivska St., Kyiv, 04119, Ukraine
gws0731512025@gmail.com

**Abstract.** The expediency of using the augmented reality in the case of using of STEM-education in Ukraine is shown. The features of the augmented reality and its classification are described. The possibilities of using the Google Expeditions and Google Lens as platforms of the augmented reality is analyzed. A comparison, analysis, synthesis, induction and deduction was carried out to study the potential of using augmented reality platforms in the educational process. Main characteristics of Google Expeditions and Google Lens are described. There determined that augmented reality tools can improve students motivation to learn and correspond to trends of STEM-education. However, there problems of using of augmented reality platforms, such as the lack of awareness of this system by teachers, the lack of guidance, the absence of the Ukrainian-language interface and responding of educational programs of the Ministry of Education and Science of Ukraine. There proposed to involve methodical and pedagogical specialists to development of methodical provision of the tools of augmented reality.

## 1 Introduction

Development of the country is depends of the education level of country. That why methods used for education of the students is provide impact on the all humanity development. There a lot of the positive educational innovation is already implemented worldwide.

## 2 Literature Review and Problem Statement

However, modern society requires the innovation methods including to increase the students motivation. The motivation factor is agreed with scientific results [9]. In the research was analyzed the motivation of students during the augmented reality (AR) education [18]. There was determined that 64.7% pupils was enjoyed the AR-education and 35.3% ware strongly enjoyed. There was no negative effect on the student's motivation.

There is described the interactive methods of education previously. This methods can be divided to the regular and periodic. The regular methods of the digital motivation of the students developed weakly in Ukraine. However, it based on the including of the interactive boards and using of the web-pages on classes. Project "The Future" [7], web-page "For the lesson" [20] and web-page of the virtual STEM-center of Junior Academy of Sciences of Ukraine (stemua.sciece) is one of most widely-used resources [15].

However, the periodic events aimed to improve motivation of students are more often used in the education system of Ukraine. However, there is no so much digitized education events. One of them is an open natural demonstration, based on the using of search skills of student that is important competence of modern people. Students are use the Google Search to research proposed question.

There is an increasing of scientific articles devoted to the AR digitalization of the classes' quantity. Thus, there was published 15 articles in the 2015, compare to 1 in 2011. The maximum articles quantity was observed in the 2014th with indicator of 18 [11].

Some scientific articles was devoted to analyze of the AR (VR)-approaches in the classes [21]. However, the aim of the article is to analyze the possibility to implement interactive augmented reality methods to the education system of Ukraine.

There was proposed to use the SamRohn 360 VR, BlakewayGigapixel, AirPano Arial Panoramaand 360 World Tours, World Tour 360, and World in 360, YouTube, Sites in 3D Virtual Tours, You Visit 360, Vatican, Kid World Citizen, Google Expeditions, Google Streets and InCell VR to use in the educational process [21].

Using of the argument reality in teaching is possibility to visualize information. Visualization of educational materials provide improving level of mastering of educational information by students [2]. Main advantage of instruments of AR is lower level of negative effects comparing to virtual reality. However, modern tendency of Ukrainian education declamation providing of STEM-approaches of educational which including using of informational technologies [24].

AR is approach to visualization of information, which include elements of virtual reality in the reality. Scheme of the augmented reality is shown on Fig. 1.

The result Lee Steven O. Zantua [27] prove that there is positive effect on the motivation of pupils. The research show that students want to learn the subject using the VR (AR). All responses based on the wants of student to use AR is positive. However, there was some problems with Google Cardboard. The most excited point of using VR (AR) for children's is possibility to "travel to the different countries" without plane. The main disadvantages are eyes hurting, the problem of lack of moving freedom and it's uncomfortable according to the children's opinion [27].

According to "The Future" project supported by Ministry of education and science of Ukraine (doc. num. № 1/9-436, 09.08.2017) AR is divided to marker, no marker, projection and VIO [7]. Lack of the possibility to exchanging methodological material is one of the limiting factors for the implementation of the AR in education. Thus, it is relevant to develop a platform for the placement of techniques and methods, in particular, the AR. The advantage of the first solution is its flexibility as one can choose any relevant combinations of the simulation environments, yet, their integration level is usually insufficient. The closed character of the second solution and its binding to a certain software platform make it relevant to be applied to solving various practical tasks and irrelevant for neural network simulation training as a network becomes a black box for a user. The fourth solution is partially platform-dependent as a neural network becomes a grey box for a user. The final solution is totally mobile and offers an opportunity to regard the model as a white box, thus making it the most relevant for initial mastering of neural network simulation methods.

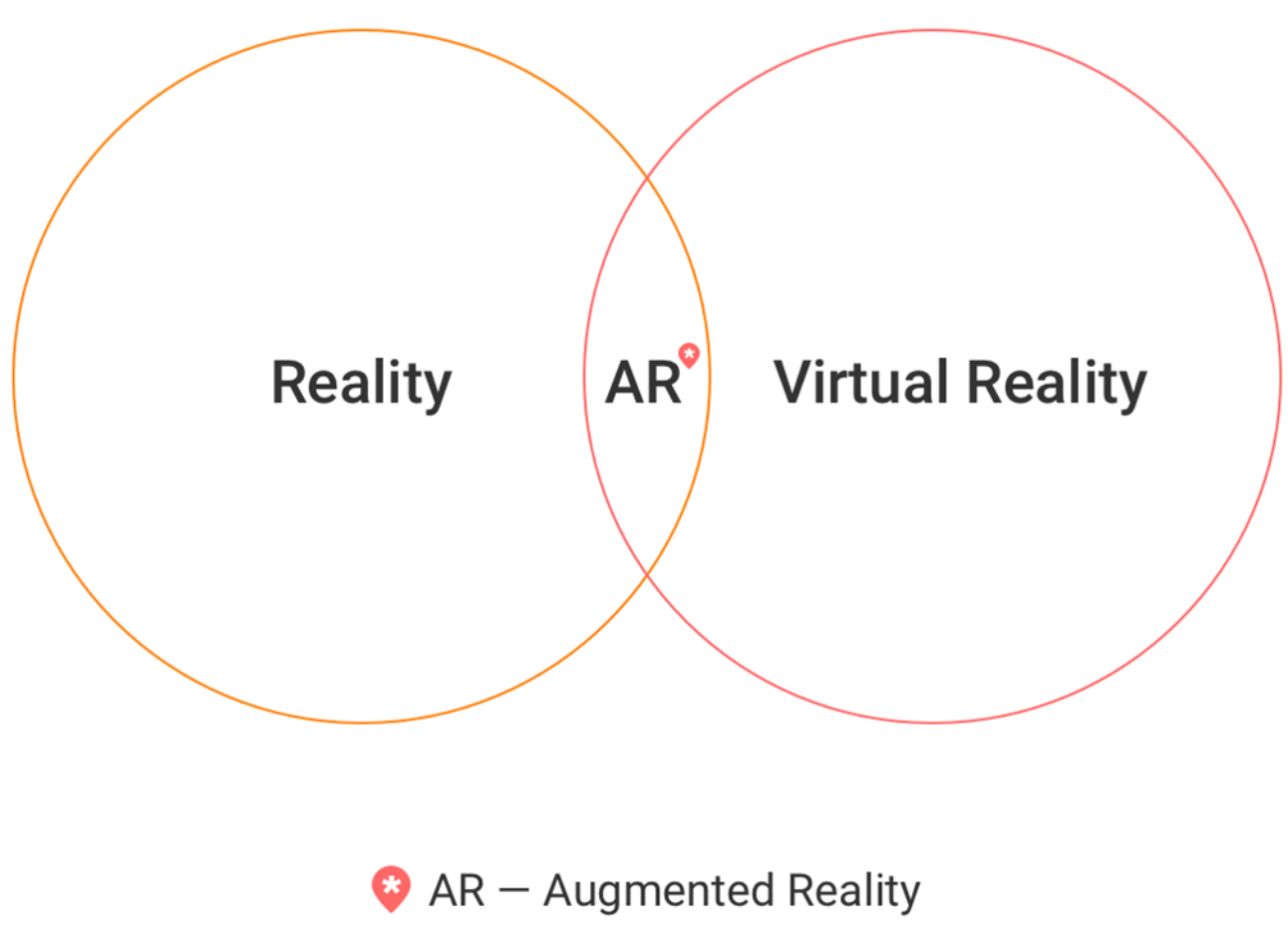

**Fig. 1.** Scheme of the augmented reality

## 3 The Aim and Objectives of the Study

The aim of the article of is to describe the possibility of using of the most relevant AR instruments for education. They are Google Expedition and Google Lens.

To accomplish the set goal, the following tasks are to be solved:

1. Learn the advanced pedagogical experience of using the Google Expedition and Google Lens in the world;
2. Compare prospects of application of using the Google Expedition and Google Lens.

## 4 Pedagogical experience of using the Google Expedition and Google Lens

Let's consider more features of functioning of both tools. Google Expedition AR is Google main instrument based on AR which imposes virtual objects on the reality fixed by phone (table) camera. This way, Google Expedition AR is include marker and projection AR. It’s already widely using in the educational systems of USA mainly in Elementary School [2; 4; 9; 27].

However, the potential of Google Expedition is not limited by using in the Elementary School. This technology can be used for visualizing of anatomy, astronomy and other environmental sciences. The example of using AR in education is present on Fig. 2.

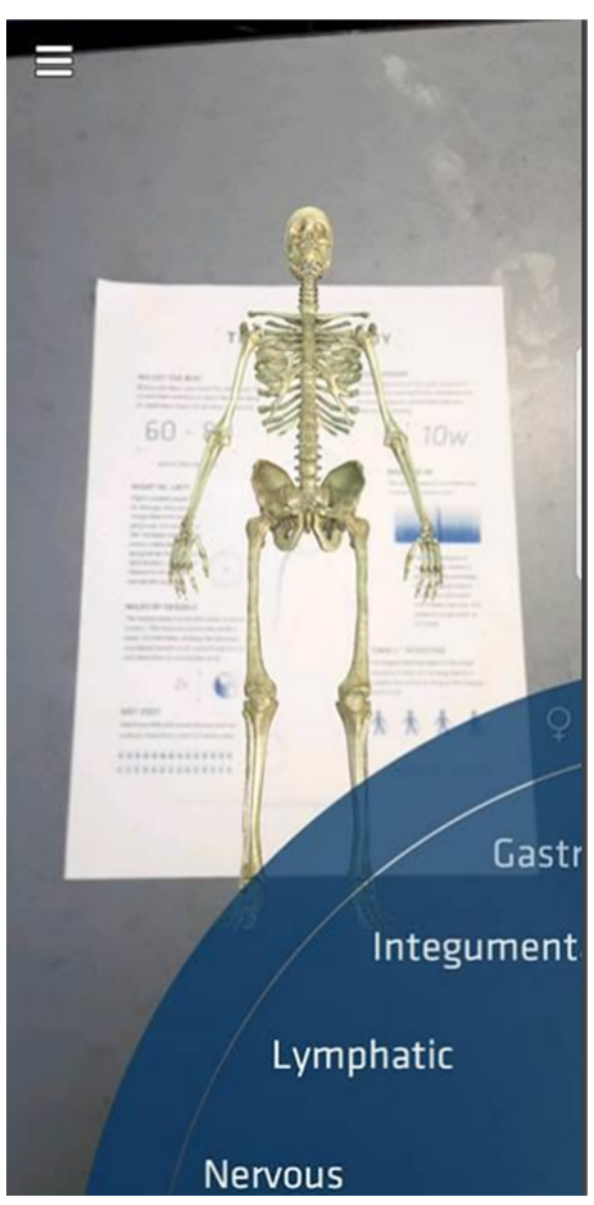

**Fig. 2.** Example of using AR in education

Google Expeditions enables teachers to bring students on virtual trips to places like museums, underwater, and outer space. Expeditions are collections of linked virtual reality (VR) content and supporting materials that can be used alongside existing curriculum.

These trips are collections of virtual reality panoramas – 360° panoramas and 3D images – annotated with details, points of interest, and questions that make them easy to integrate into curriculum already used in schools.

Google is working with a number of partners, including: WNET, PBS, Houghton Mifflin Harcourt, the American Museum of Natural History, the Planetary Society, David Attenborough with production company Alchemy VR and many of the Google Arts & Culture museum partners to create custom educational content that spans the universe. With over 300 Expeditions currently available, the content touches on a wide range of subjects including historical landmarks, natural wonders, college campuses, careers and more [1].

Google Expeditions has already serviced over 2 million students [17]. Expanding into augmented reality will allow teachers to further engage their students by not only letting them view the subjects from every angle, but by letting them interact and share the experience with each other as they do. Allowing students to gather around an object or scene more closely mirrors an open classroom environment, having significant benefits of each student holding a Google Cardboard over their eyes. Manual of using Google Expedition during lessons are actively developed by progressive teachers from the UK and the United States [25].

Using of Google Expedition may improve the size understanding of the natural phenomena. For example, the anatomy learning can be upgraded this case due to possibility of size understanding. Studying human anatomy with virtual reality gives students a better grasp of the relative size of the different organs and parts which is better to memorize it [4].

In the contrast of widely used and well-known Google Expeditions AR's applications, there exist Google Lens which is innovated of education. The technology was presented in 2017 on the Google I/O conference [8]. Lens recognizes the image and its show the relevant information about it. The approach is integrated to the Google Photo and Google Assistant. This instrument use Artificial Intelligence for the best work and it self-development. This way, Google Lens is concerns to the marker AR.

So, the main potential of the instrument is to learn the historical and cultural features under traveling. However, our opinion that there is much bigger potential to use it for the biological and other environmental researches.

The approach is simplify the research process. The example of Google Lens using in the biological research is present on the Fig. 3.

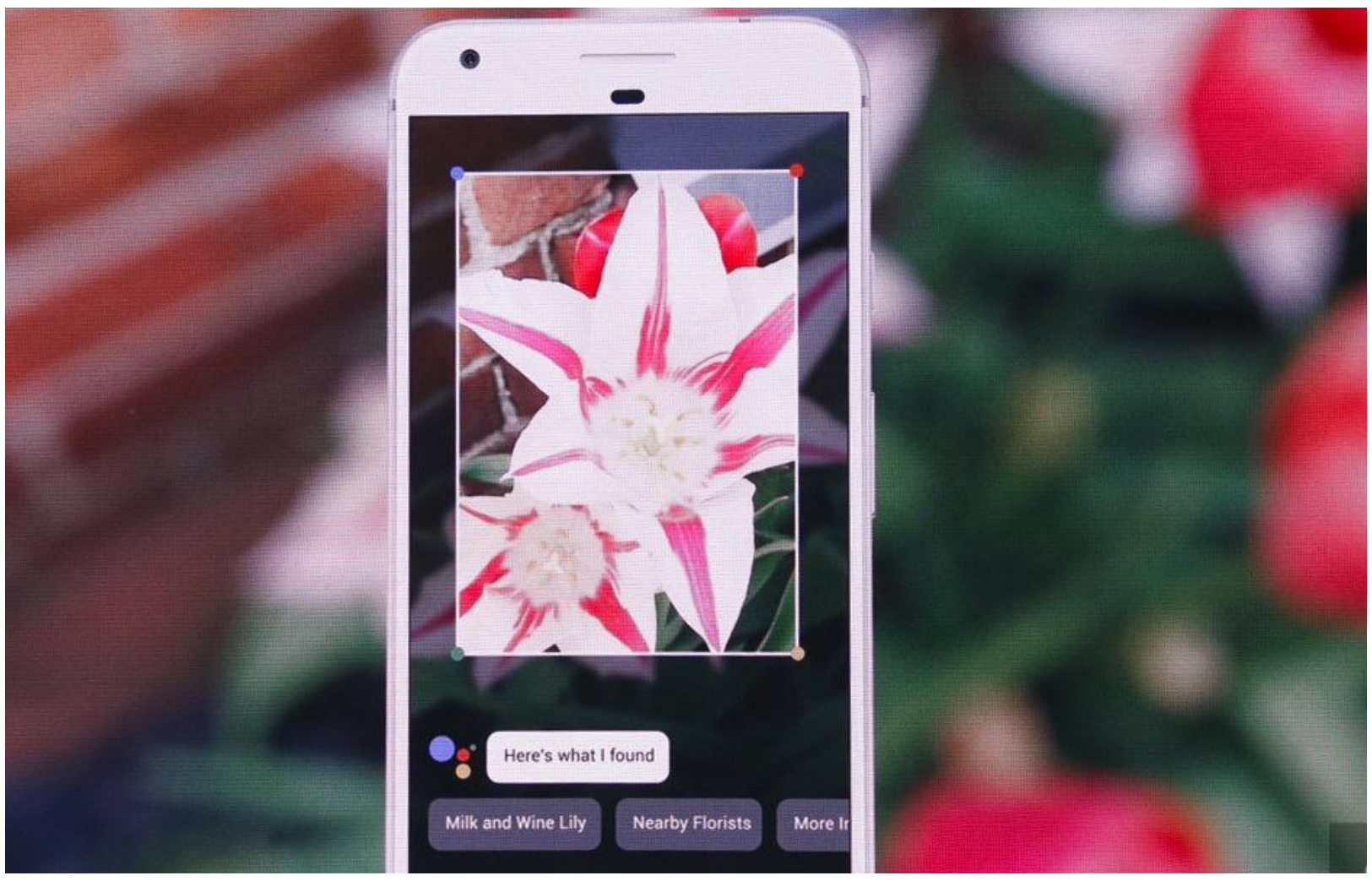

**Fig. 3.** The example of Google Lens using in the biological research

The instruments can be used on the classes and during self-educational process. The studying programs of Ukraine is recite including of the research works based on the research method. The research method consist of background, construction, research and analysis part. Detailed information of the research method is described in our previous works and sciencebuddies.org web-page. One of the limiting parts of providing the research work is actual research process. The problem of providing the research is lack of equipment. Both approaches described previously can simplifies the requirements for the research process due to possibility personal self-phone using.

However, increasing of the student's motivation is one of the problem of the modern education. Nowadays there is digitization process which led to concentration of the children attention to the digital visualized information and that why there is the problem of the motivation of students. Both, Lens and Expedition can improve the motivation level of students due to its interactivity. However there is noted that the main advantage of augmented reality is simplification of the linking between other people and interact with their surroundings [16] and provide improving of the teamwork. The important AR advantage is that role of teacher is changed to the facilitator's role who helps the students explore and learn and its increase enjoy of the education due to the students control learning process [3; 6; 26]. The pupils are concentrated on the education process ignoring the distractions [10; 11].

Studding of the low-spatial ability learners is facilitates due the lack of the extraneous cognitive overloading [14]. The AR can simulate the dangerous situations to be ready for them [22].

However, the AR is important component of the STEM-education. The engineering aspect can be provided by possibility to test the prototype in the virtual environment before create it [23]. The science aspect can be achieved by the testing of the theories and hypotheses in the virtual environment [12]. The main positive aspects of the represented methods are presented in the Table 1.

**Table 1.** The main positive aspects of the represented methods

| | **Google Expedition** | **Google Lens** |
|---|---|---|
| *Motivation* | Due to IT using, and it's interactivity | Provided by possibility to use personal phones any time to research |
| *Interactivity* | Provided by possibility of the Google Expedition to visualize natural phenomena | Interaction with any objects |
| *Knowledge increasing* | Due to the information visualization | Due to the possibility of research any object any time |
| *Other advantages* | Simplification of the linking between other people and interact with their surroundings, improving of the teamwork, introduction of facilitator's role, possibility to study of the low-spatial ability learners, size understanding, better memorizing, possibility to simulate the dangerous situations, providing STEM-education | |

Using of these AR approaches depends on the few factors. The potential analysis of implementation of Google Lens and Google Expedition AR is presented in the Table 2.

The AR-equipment can be devoted to chip and expensive. The simplest example of the simple AR-equipment is Google Cardboard with cost of 15$. However, there exist

the Google daydream with 99$ of cost, Microsoft HoloLens with cost of 3000$ and google glasses with cost of 800$.Thus, the STEMUA platform allows teachers to develop methodological material and deposit it to the platform. Methodical materials are automatically systematized in the database of the platform, and the materials are foreseen mainly in Ukrainian, which meets the requirements of the Ministry of Education and Science. Consequently, the platform is able to meet the teachers' methodological needs regarding the use of AR in the classes.

**Table 2.** The analysis of implementation of Google Lens and Google Expedition AR

| | **Google Expedition** | **Google Lens** |
|---|---|---|
| *Abstract* | AR instrument | Image analyzing system |
| *Approaches in education* | Physics, chemistry, biology, geography, history, architecture | Biology, history, architecture, mineralogy, geology, engineering |
| *Pedagogical aspects* | Lack of teachers awareness of this instruments, lack of the methodical achievements and there absence of Ministry of science and education recommendation about it | |
| *Technical problems* | There is no official Google office in Ukraine (or it's weak communication) and lack of equipment | High equipment cost of the Lens supported stuff, there some mistakes of under working |
| *Other problems* | There is necessary to be careful with AR-devices due possibility of damage the device during which might damage electronic components, problems GPS errors effect on the accurate of markerless AR programs | |

## 5 Conclusions

1. Google Lens and Google Expedition can enhance students' motivation to learn and correspond to trends in STEM education.
2. The use of these tools is limited by a number of factors, such as the lack of knowledge of this system from teachers, the lack of guidance on the use of this system, the absence of prevailing the majority of the Ukrainian-language interface and the absence of stamp Ministry of Education and Science of Ukraine.
3. The indicated problems can be solved by involving methodical-pedagogical workers in the development of methodical provision of the instruments of the complemented reality.
4. We offer to provide classes with short-time using of the Google Expeditions and Google Lens due to its advantages kind simplification of the linking between other people and interact with their surroundings, improving of the teamwork, introduction of facilitator's role, possibility to study of the low-spatial ability learners, size understanding, better memorizing, possibility to simulate the dangerous situations, providing STEM-education.
5. Long time using can affect negatively due to the yes hurting, the problem of lack of moving freedom and it's uncomfortable.